\documentclass[twocolumn,showpacs,preprintnumbers,amsmath,amssymb]{revtex4-1}

\usepackage{graphicx} \usepackage{amssymb} \usepackage{hyperref} \textwidth 17.5cm
\begin{document}

\title{Low temperature thermomagnetic properties of very heavy fermions suitable for
adiabatic demagnetization refrigeration}

\author{Julian G. Sereni} 

\address{Physics Department, CAB-CNEA, CONICET, 8400 San Carlos de Bariloche, Argentina}
\begin{abstract}

{With the aim of improving the performance of classical paramagnetic salts for adiabatic
refrigeration processes at the sub-Kelvin range, relevant thermodynamic parameters of some
new Yb-based intermetallic compounds are analyzed and compared. Two alternative potential
applications are recognized, like those requiring fixed temperature reference points to be
reached applying low intensity magnetic fields and those requiring controlled thermal
drift for temperature dependent studies. Different thermomagnetic entropy $S(T,B)$
trajectories were identified depending on respective specific heat behaviors at very low
temperature. To gain insight into the criteria to be used for a proper choice of suitable
materials in respective applications, some simple relationships are proposed to facilitate
a comparative description of their magnetocaloric behavior, including the referent
Cerium-Magnesium-Nitride (CMN) salt in these comparisons.}

\end{abstract} \date{\today} \maketitle

\section{Introduction} 

\begin{figure}[tb] \begin{center} \includegraphics[width=20pc]{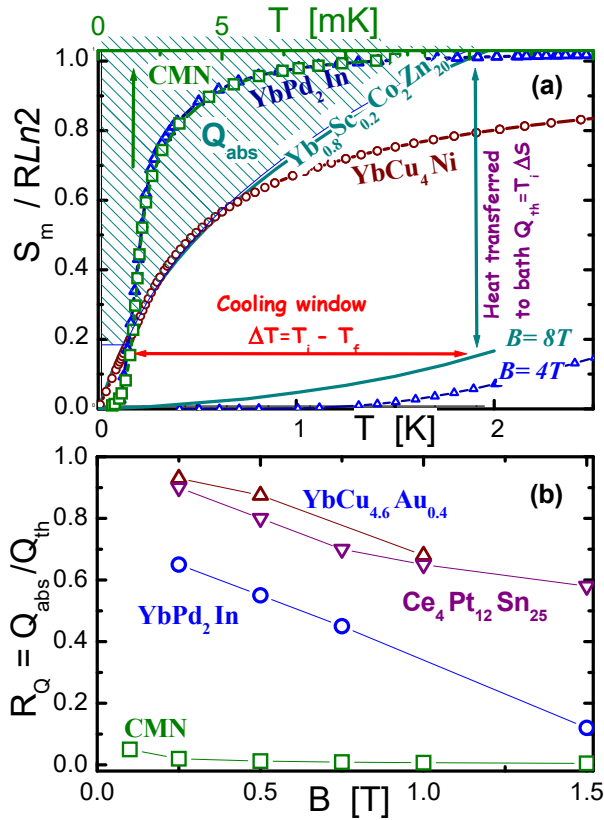}
\end{center} \caption{(a) Exemplary $S_m(T)$ trajectories for three different types of
behaviors \cite{YbPd2In,Gegenw,YbCu4Ni}. CMN salt (upper $T$ scale) is included for
comparison \cite{CMN1}. Vertical arrow: heat transferred to the bath
$Q_{th}$. Horizontal arrow: cooling window $\Delta T$. Shadowed triangle (top left)
represents the heat absorbed $Q_{abs}$ by these materials in the process of thermal
stabilization taking as reference the (Yb,Sc)Co$_2$Zn$_{20}$ system \cite{Gegenw}.
(b) Heats ratio $R_Q = Q_{abs}/Q_{th}$ as a function of magnetic field evaluated from $T_i =1$\,K. }
\label{F1} \end{figure}

Technical requirements imposed by photo-detectors installed in orbital satellites
\cite{cryogen} and the shortage of He$^4$/ He$^3$ gases due to an increasing demand and
reducing supply \cite{He3He4} have powered the search of new materials able to improve the
cryo-performance of classical paramagnetic salts \cite{CMN1} for adiabatic demagnetization
refrigeration (ADR) processes \cite{Pobell}. These novel demands require cryo-materials
suitable to operate under extreme conditions at the sub-Kelvin range of temperature. Two
typical demands can be mentioned, one of them requiring a thermally stable cryo-source tuned at a
precise operating temperature, i.e. acting as fixed thermal point (FTP). The other
corresponds to provide a controlled thermal drift (CTD), maximizing the heating absorption
in an extended range of temperature.

Apart from these specific potential needs, some common demands arise like: to
have 'friendly' synthesis, good chemical stability, high thermal conductivity, low eddy
currents and large volumetric entropy. Moreover, to have good performance at low applied
field together with mechanical hardiness and to support dehydration \cite{Pobell} under
vacuum conditions are 'sine qua non' conditions for satellite applications.

As it can be seen in  Fig.~\ref{F1}a, the ADR thermodynamic cycle can be described by three 
successive steps. The first corresponds  to the application of
magnetic field in isothermic conditions with a heat transference $Q_{th}$  to the thermal bath at the 
initial temperature $T_i$.  In this branch, depicted by a vertical arrow, the heat given 
out depends on the entropy reduction induced by increasing the applied field as: $Q_{th} =T_i \Delta 
S_m(B)|_{T_i}$. 

The following step consists in the adiabatic cooling process where the cooling window 
$\Delta T= T_i - T_f$ (with $T_f$= the final temperatures) is also included in Fig.~\ref{F1}a as 
an horizontal arrow where the thermodynamic equation: $dS(T,B) = (\partial S/\partial T)|_B dT + 
(\partial S/\partial B)|_T dB = 0$ applies.  This formula can be written including measurable   
parameters as $dS(T,B) = C_m/T dT +
 \partial M/\partial T dB = 0$ because $\partial S/\partial T = C_m /T$ and $\partial S/\partial B|_T = 
 \partial M/\partial T|_B$.
 
 \begin{figure}[tb] \begin{center} \includegraphics[width=21pc]{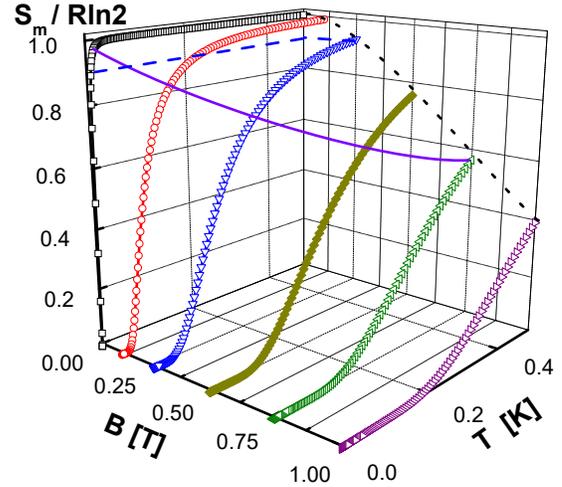}
\end{center} \caption{Exemplary $S_m(T,B)$ trajectories at different fields for CMN, after
\cite{CMN1} in a 3D representation. Continuous curve indicates the maximum curvature of
the $S_m(T,B)$ trajectory, dashed line the thermal window $\Delta T = T_i -T_f$ for $B =
0.25$\,T and dotted curve the $\partial S_m/\partial B|_{T_i}$ variation at $T_i=0.4$\,K.}
\label{F2} \end{figure}

Along the third step the cryo-material is devoted to stabilize or control the temperature drift by 
absorbing heat $Q_{abs}$. This heat is represented  in Fig.~\ref{F1}a by the shadowed area: 
$Q_{abs} = \int_{S_i}^{S_f} T dS_m$. Since this parameter depends on the heat capacity along
the temperature range, it corresponds to the enthalpy variation at $B=0$ because
$T\partial S_m = C_m \partial T$. From this expression it becomes clear that materials without magnetic order but with robust magnetic moments are relevant for ARD purposes. These two conditions, that look contradictory, are realized in some recently discovered materials.   

Since $C_m/T$ is proportional to the density of excitations, the rich family of
intermetallic heavy Fermions (HF) provides asignificant number of candidates for ADR
applications. Among them are the recently synthesized Yb-intermetallic compounds
\cite{YbCu5xAux,YbPd2In,YbCu4Ni,Gegenw,YbPt2Sn,JangNatur}, some of which have been tested
as proper candidates for the proposed improvements because of their very large
$C_m/T|_{T\to 0}$ values. These compounds were labeled as very HF (VHF) \cite{JLTP18}
because they largely exceed the values of classical HF \cite{Stewart01}. According to
their physical behavior within the mK range three well defined types of $C_m(T)/T$
behaviors were recognized \cite{SerPhysB18}: (I) the actual VHF with $C_m/T|_{T\to 0} >
10$\,J/molK$^2$, (II) those showing a constant $C_m/T|_{T\to 0}$ 'plateau' around 7\,J/molK$^2$ below a characteristic temperature $T^*$ and (III) those belonging to the
non-Fermi-liquid (NFL) family \cite{Stewart01} because of their $C_m(T)/T \propto -
\ln(T/T_0)$ dependence.

All these paramagnetic materials do not order magnetically down to about 200\,mK despite
of their robust magnetic moments, providing large $\partial M/\partial T$ coefficients. As
mentioned before, the entropy computed as: $S_m(T) = \int C_m/T dT$ is the basic parameter
for the study of ARD processes because its thermal trajectory characterizes the efficiency
and applicability of each cryo-material at the 'sub-Kelvin' range.

The scope of this work focuses mainly on the analysis and comparison of the thermomagnetic properties of new intermetallic compounds exhibiting different behaviors which can be applied in alternative applications.
In the search of a practical criteria to identify the mot appropriated material for each application, some simple relationships are proposed to facilitate a
comparative description of their magnetocaloric behavior. The well known salt
Cerium-Magnesium-Nitride (CMN), largely used for ADR processes because it remains paramagnetic down to 2\,mK \cite{CMN1,CMN2}, is also included in this study as a reference to classical salts.

\section{Analysis of thermomagnetic properties}

As expected, FTP and CTD applications require different types of $S_m(T)$ trajectories to
optimize respective ADR processes. In Fig.~\ref{F1}a, three exemplary types of trajectories are
depicted showing that they can be sorted between those suitable as FTP providers and two
alternatives of moderate increase of $S_m(T)$ more indicated for CTD operation conditions.

In the figure one may appreciate how the $S_m(T)$ trajectory of CMN matches with that of YbPd
$_2$In by simply normalizing respective scales of temperature. Therefore this reference salt can be  included in group (I) like a member of the YbT$_2$X family (where T=Pd and Pt and X=In and Sn). 
These compounds are better indicated for FTP processes because of their steep $\partial S/ \partial T
 = C_m/T$ slope allows to tune them within a relative narrow thermal window around the required 
working range, e.g. according to satellite photon-detectors requirements. Differently,YbCu$_4$Ni 
and (Yb,Sc)Co$_2$Zn$_{20}$, also included in  Fig.~\ref{F1}a belong to groups (II) and (III) 
respectively and better apply for CTD requirements because of their moderate slope allows a more 
convenient 
heat absorption more homogeneously distributed in temperature. Within group (II) one can mention 
YbCu$_4$Ni \cite{YbCu4Ni} which represents the family of compounds showing a $C_m/T|_{T\to 0}\approx 7.5\pm 0.5$\,J/mol\,K$^2$ 'plateau' followed by a common power law above a change of
regime at $T=T^*$ \cite{JLTP18}. The compounds included in group (III) are characterized by 
showing NFL behavior, like (Yb,Sc)Co$_2$Zn$_{20}$ \cite{Gegenw} and CeNi$_{9-x}$T$_x$ (where T = Co \cite{Ni9Cox} with $x = 0.1$, and T = Cu with $x = 0.4$ \cite{Ni9Cux} respectively).

One of the basic questions to be addressed in the ADR planning is the choice of the more
appropriated material for each specific application. Since the number of involved parameters provides  
a rich spectrum of possibilities for an optimal application, the identification of physical quantities for a 
convenient comparison between different systems is advisable. Some proposition can be already 
extracted from  Fig.~\ref{F1}a involving the mentioned heat transfers involved in the thermodynamic 
cycle: $ Q_{th}$ and $Q_{abs}$. A sort a efficiency in the heat management can be defined by the 
capability of the system to absorb heat in confront to the amount previously transferred to the bathat 
different fields through the ratio $R_Q(B)= Q_{abs}(B)/ Q_{th}(B)$.  In Fig.~\ref{F1}b this 
comparison is presented for four representative compounds of different types of $S_m(T,B)$ trajectories whose data are available in the literature as a function of applied fields within the $B\leq 1.5$\,T range. For practical reasons this comparison was done taking $T_i =1$\,K, with the magnetic field within an eventual satellite application range.    

This comparison includes the $R_Q(B)$ ratio obtained from compounds belonging to groups (I): CMN, YbPd$_2$In and Ce$_4$Pt$_{12}$Sn$_{25}$, and group (II): YbCu$_{4.6}$Au$_{0.4}$.
There, one may appreciate that a better ratio is provided by a system with lower $\partial S/\partial T$ slope (like YbCu$_{4.6}$Au$_{0.4}$) than by one with a larger slope (like CMN). An intermediate
alternative is provided by the VHF YbPd$_2$In. From this first comparison one learns that the compound reaching the largest $C_m/T$ value at  very low temperature is not the best for the heat absorption in a CDT process. Nevertheless, their  $Q_{abs}$ values can be considerably increased by tuning $T_f$ at higher values by not reducing the field down to zero.

\begin{figure}[tb] \begin{center} \includegraphics[width=20pc]{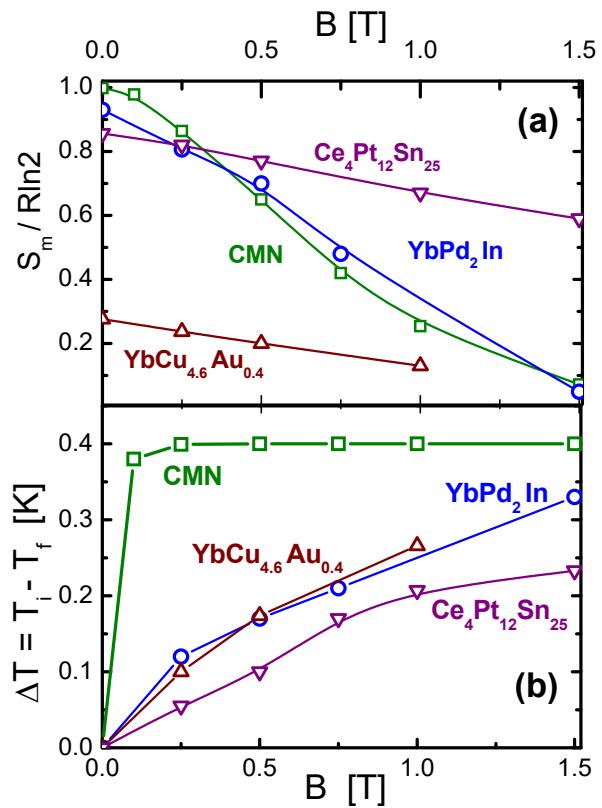} \end{center}
\caption{Comparison of two characteristic parameters between CMN \cite{CMN1}, YbPd$_2$in
\cite{YbPd2In},YbCu$_{4.6}$Au$_{0.4}$ \cite{YbCu4Ni} and  Ce$_4$Pt$_{12}$Sn$_{25}$ \cite{Ce4}. 
as a function of applied field:
a) Field dependence of the entropy $S_m(B)$ at 0.4\.K and b) cooling window $\Delta T$ at different fields in  the adiabatic process.} \label{F3} \end{figure}

Before to proceed for further comparisons it is convenient to analyze the role of the involved 
parameters into the different $S_m(T,B)$ trajectories. For such a purpose one can use the 
properties of the well known CMN salt available in the literature \cite{CMN1,CMN2}, resumed in a 
$S_m(T,B)$ diagram in Fig.~\ref{F2}. In that figure it is evident that the entropy decrease, indicated as a doted curve at $T_i$, has a maximum slope at some intermediate value. This means that an optimum  $\partial S_m/\partial B|_{T_i}$ variation occurs at 
$k_B T=\mu_B B$ because thermal and Zeeman splitting become comparable. This condition can be traced in the 3D figure by following the maximum curvature of the $S_m(T,B)$ trajectory along different fields, indicated as a continuous curve in the figure.

Another characteristic is related with the size of the cooling window as a function of the applied field, 
as a sort of magneto-cooling efficiency: $\Delta T/\Delta B$. In the case of CMN one can see that 
$\Delta T$ already reaches 90\% of its maximum value at around $B = 0.25$\,T as it is
indicated by a dashed line in  Fig.~\ref{F2}. Therefore, further increase of field does not
improve the cooling window. In this case, the initial temperature is taken as $T_i=0.4$\,K in order to present this feature in a convenient energy scale.

Both quantities are compared for different intermetallic compounds in Fig.~\ref{F3} as a function of 
field. The different $S_m(B)$ trajectories presented in Fig.~\ref{F3}a clearly indicate that CMN and 
YbPd$_2$In are the most sensitive under applied magnetic field. Notably the low sensitivity of YbCu
$_{4.6}$Au$_{0.4}$ and Ce$_4$Pt$_{12}$Sn$_{25}$ is due to opposite reasons, the former because 
its ground state shows reminiscence of a liquid of spins\cite {YbCu5xAux} while the later because of its antiferromagnetic character \cite{Ce4}.

The field dependence of the cooling window $\Delta T$ is collected in  Fig.~\ref{F3}b. As expected, 
CMN reaches its maximum value at very low field because its ordering temperature is $T_N = 2$\,mK 
whereas for intermetallic compounds  $\Delta T$ increases gradually because their characteristic 
temperatures are quite comparable with the applied field.  

From these comparisons, one concludes that the systems suitable for temperature
stabilization (FTP) have lower capacity to control thermal drift but they are more sensitive to magnetic. The contrary occurs for those with higher capacity to absorb heat within a
more extended range of temperature (CTD).

\section{Thermal properties of VHF}

\begin{figure}[tb] \begin{center} \includegraphics[width=20pc]{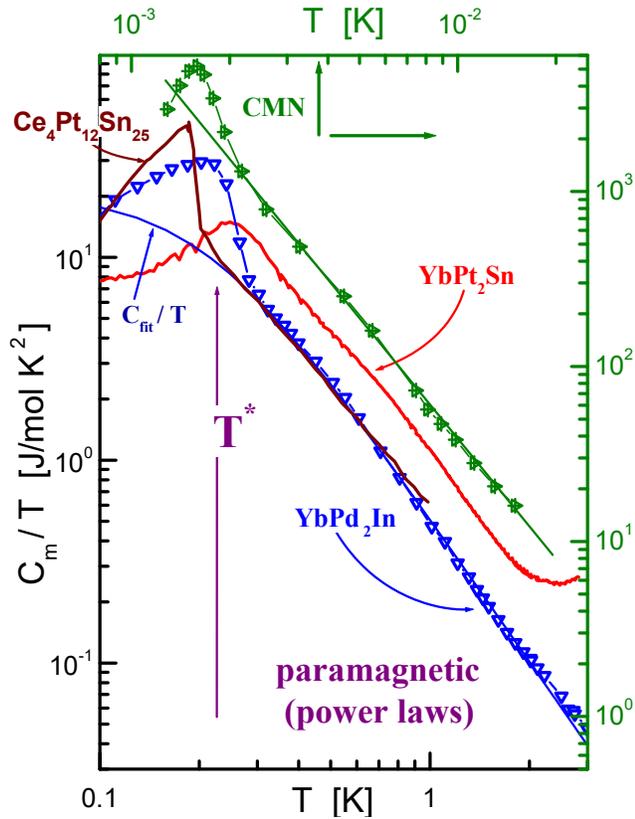}
\end{center} \caption{Low temperature specific heat of some VHF in a double logarithmic
scale. Left and lower axes: YbPt$_2$Sn \cite{YbPt2Sn}, YbPd$_2$In \cite{YbPd2In} and
Ce$_4$Pt$_{12}$Sn$_{25}$ \cite{Ce4} with similar power law dependence (left and lower
axes). Dashed curve represents the fit at $T>T^*$ with $C_{fit}/T = 0.5/( T^{2.35}+0.02)$.
Right and upper axes: CMN is included (after ref. \cite{CMN1}) for comparison and discussion,
with the straight line representing a $T^{-2.3}$ dependence.} \label{F4} \end{figure}

Turning our attention to the thermal properties of the VHF-YbT$_2$X family of group (I),
collected in Fig.~\ref{F4}, one notices that they show the highest $C_m/T|_{T\to 0}$ ratio
among Yb-intermetallics, reaching a maximum of $\approx 30$\,J/molK$^2$ at $T^* \approx 200$\,mK after subtracting the Yb-nuclear contribution \cite{YbCu5xAux}. Among Ce-based
intermetallic compounds, to our knowledge only one can be included into this group. These
values are obviously exceeded by the Ce paramagnetic salt CMN. It is worth noting that
these large $C_m/T|_{T \to 0}$ values become quite similar if respective $C_m(T)$
dependencies are compared as it will be done in the following Section.

A common feature of their $C_m(T)/T$ increase at $T>T^*$ is the power law thermal
 dependencies: $C_{fit}/T = G/(T^D+T_0^D)$ \cite{Ser07} within the paramagnetic range, see 
 for example the dashed curve in Fig.~\ref{F4}. They exhibit similar exponents: $2.27 \leq D \leq
2.35$ and $T_0$ values: $0 \leq T_0 \leq 0.1$. In these compounds the slight
deviation from a pure power law dependence is accounted by $T_0$ that can be attributed
to a reminiscence of a very weak Kondo-type interaction. As expected, CMN shows a pure
power law dependence because of the lack of conduction electrons.

\begin{figure}[tb] \begin{center} \includegraphics[width=20pc]{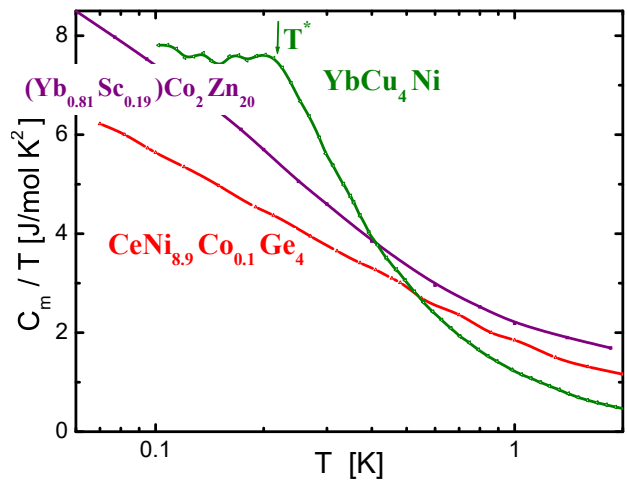}
\end{center} \caption{Specific heat temperature dependence of some selected compounds from
groups (II) and (III) in a semi-logarithmic representation. $T^*$ marks the lower limit of
the power law dependence for YbCu$_4$Ni.} \label{F5} \end{figure}

For comparison with the members of other groups, the $C_m(T)/T$ of some representative
compounds of group (II): YbCu$_4$Ni \cite{YbCu4Ni}, and group (III):
(Yb,Sc)Co$_2$Zn$_{20}$ \cite{Gegenw} and CeNi$_{8.9}$Co$_{0.1}$ \cite{Ni9Cox} are
collected in Fig.~\ref{F5} in a logarithmic $T$ dependence. There one can see how 
the density of excitations is distributed in energy in NFL systems (with a $C_m(T)/T \propto
- \ln(T/T_0)$ dependence) respect to those with a power law ($C_{fit}(T)/T \propto
1/T^{1.24}$ \cite{YbCu4Ni}) like YbCu$_4$Ni at $T>T^*$ also included into the figures.

\section{Magnetic interactions and frustration}

The absence of long range order is a 'sine qua non' condition for a cryo-material because
it allows to keep large volume entropy (i.e. entropy per molar volume) down to very low
temperature. Two main reasons may allow to retain the paramagnetic GS despite of the
robust character of their magnetic moments: very weak magnetic exchange between
neighboring spins mostly observed in group (I) \cite{YbPt2Sn} or magnetic frustration usually
present in
group (II) \cite{YbCu4Ni}. A third possibility occurs in the proximity to a quantum critical point that
is related to the NFL behavior characterizing group (III) \cite{Gegenw}.

YbT$_2$X compounds are the good representatives of weak magnetic exchange: $\aleph =
J_{i,j} \times S_iS_j$ because the usual RKKY interaction in intermetallics depends on the
conduction density of states: $\delta(\epsilon_F)$, becuase $J_{i,j} = J_{loc}
\delta(\epsilon_F) f(1/d^3)$. The large residual resistivity of these Yb-based compounds,
coincident with a small $\partial \rho/\partial T$ coefficient \cite{YbPt2Sn} and very low
Sommerfeld coefficient $\gamma \approx \delta(\epsilon_F)$ observed in the non-magnetic
reference compounds LuPt$_2$In/Sn converge to very low $\delta(\epsilon_F)$ values. In the
CMN salt a RKKY interaction is certainly excluded because $\delta(\epsilon_F)= 0$ and the
expected dipolar interaction between quite distant Ce atoms decays exponentially.

The 3D geometrical frustration, favored by the pyrochlore structure of many components of
group (II), may guarantee the lack of long range magnetic order but not the largest $C_m/T$
values (see e.g. \cite{SerPhysB18}). Although the compounds of the NFL group (III) show more
extended distribution of magnetic excitations, favoring the $R_Q(B)$ ratio, their proximity
to a quantum critical point is related with a weakening of their magnetic moments with the
consequent reduction of the $\partial M/ \partial T = \partial S/\partial H$ slopes.

\begin{figure}[tb] \begin{center} \includegraphics[width=20pc]{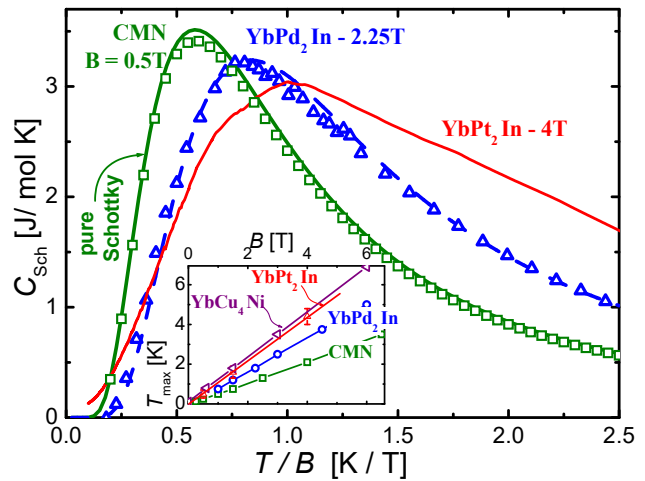}
\end{center} \caption{a) Selected Schottky type anomalies of $C_m(T)$ scaled with
respective applied fields. Inset: linear field dependence of the temperature of
corresponding $C_m(T,B)$ maxima.} \label{F6} \end{figure}

The maximum in $C_m(T)/T$, labelled as $T^*$ in Fig.~\ref{F4}, may be regarded as a
standard second order phase transition induced by long range magnetic interactions.
However, in these systems more than 40\% of the magnetic entropy is already condensed
above $T^*$ and the divergent character of the $C_m(T>T^*)/T$ power law dependence points
more likely to a so-called 'entropy bottleneck' effect \cite{Sereni15} caused by the
thermodynamic constraint in the entropy evolution imposed by the Nernst postulate:
$S_m|_{T\to 0}\geq 0$ \cite{Pippard}. This constraint induces relevant modifications in
the entropy trajectory because it imposes a change of the free energy minimum. Depending
of the nature of the alternative munimum (at lower temperature) the system may slide into different types of ordered phases that react differently under the application of magnetic field.
Ce$_4$Pt$_{12}$Sn$_{25}$ for example \cite{Ce4} shows a typical antiferromagnetic
behavior, whereas YbPd$_2$In tends to a critical pint at $B\approx 0.75$\,T
\cite{YbPd2In}. Except for the Ce salt, where the magnetic interaction seems to be
quite strong (and of dipole-dipole nature), the other systems with weak interactions show a similar behavior among them, transforming their specific heat jumps into the Schottky type anomalies once the external field has overcome the weak internal (molecular) field of
the order of $\mu_B B \approx k_BT^*$. Some examples are collected in Fig.~\ref{F6}, where one may also note the similar values of their $C_m(T)$ that were mentioned in the previous section. 

\section{Magnetic field effects on the ground state}

The curves included in Fig.~\ref{F6} are compared using a normalized $T/B$ abscissa selecting field that overcome the weak interactions governing the zero field ground state which cause the
specific heat jumps observed in Fig.~\ref{F4}. These thermal characteristics imply that in the analyzed VHF compounds there is an
equivalent field induced Zeeman splitting effect on the GS doublet. Furthermore, the
observed scaling on '$T/B$' at $B> 0$ confirms the weak nature of the interactions present
at $B=0$ because the $C_m(T)$ tendency to a Schottky type anomaly indicates that the ground state doublet behaves as that of a non interacting (or single) particle. 

Notably, the $C_m(T)$ of CMN kindly fits into a pure Schottky anomaly for two
'Dirac' levels with the expected $C_m(T_{max}) = 3.6$\,J/molK \cite{Elsevier}, whereas the
two YbT$_2$In intermetallics show a slightly broadened anomaly with a concomitant reduced
$C_m(T_{max})$ value. This difference is likely due to the nature of the interctions
dominating the GS at $B=0$. Taking profit that the temperature of respective maxima $T_{max}(B)$ 
increases quite linearly with field, the giromagnetic factors of respective ground state doublets     
$g_{GS}$ can be evaluated from: $T_{max}(B) = 0.42 *g_{eff} * B* (\mu_B/k_B)$ \cite{YbSn3Ru4}.
These field dependencies are compared in the inset of Fig.~\ref{F6} as a $T_{max}(B)$
dependence. One should remark that no Schottky-like behavior is allowed down to $T\to 0$ because 
of Nernst postulate, which compels the system to reduce the two fold degeneracy of its GS
through any alternative minimum within the free energy surface, like e.g. the transitions presented in 
Fig.~\ref{F4}. 

\section*{Summary}

From this analysis one concludes that different types of applications can be distinguished
within the sub-Kelvin range of temperature, recognized as those able to provide 'fix
thermal points - FTP' (e.g. for photon-detectors) and those for a 'controlled thermal
drift - CTD' (e.g. for standard cryostats). Since respective applications can be optimized
by a proper choice of cryo-materials that show different types of entropy trajectories, a
comparative analysis using simple characterizing parameters is proposed. One finds that,
in the case of photon-detectors working around the 100\,mK range, they can be thermally
stabilized using some recently discovered VHF compounds as FTP materials. The analysis of
their cooling efficiency, $\Delta T$ vs. $\Delta B$, indicates that 90\% of the thermal window
is already reached at relatively low fields ($B\leq 1$\,T). This family of Yb-based
compounds from group (I) show high volume entropy and convenient
sensitivity to magnetic field variation because of a large $\partial M/\partial T$.
The field dependence of their paramagnetic GS can be simply described using a two level
Schottky anomaly for the specific heat once the applied field has overcome their weak GS interactions. 
Complementary, the compounds of groups (II) and (III), with softer $C_m(T)/T$ dependencies show
higher capacity for heat absorption $Q_{abs}$ distributed in a larger range of temperature as 
required for CTD applications. Although they are not so sensitive to magnetic field variation, one may
conclude that there is not a unique 'best' system for ADR but different optimal choices 
depending on the type and range of application. In this case one may remark that paramagnetic salts 
do not offer suitable compounds for CTD as provided by intermetallic ones.

From the analysis of the physical phenomena occurring in the range of energy where quantum effects  
become dominant one learns that, while quantum mechanisms intervenes in the formation of 
alternative GS, the Nernst postulate imposes conditions on their thermodynamic feasibility as $\to 0$.  
This means that quantum mechanics and thermodynamic laws work entangled providing and defining 
the proper ground state.


\begin{thebibliography}{00}

\bibitem{cryogen} B. Collaudin and N. Rando, in {\it Cryogenics in space: a review of the
missions and of the technologies}, Cryogenics {\bf 40} (2000) 797.

\bibitem{He3He4} R.T. Kouzes and J.H. Ely, in {\it Status summary of He$^3$ and neutron
detection alternatives for homeland security} Report PNNL- 19360, Pacific Northwest
National Laboratory, 2010.

\bibitem{CMN1} W.F. Giauque, R.A. Fisher, E.W. Hornung, G.E. Brodale, in {\it
Magnetothennodynamics of Ce$_2$Mg$_3$(NO$_3$)$_{12}$.24H$_2$0}, J. Chem. Phys. {\bf 58}
(1973) 2621.

\bibitem{Pobell} F. Pobell, in {\it Matter and Methods at Low Temperatures},
Springer-Verlag (1991).

\bibitem{YbCu5xAux} I. Curl\`ik, M. Giovannini, J.G. Sereni, S. Gabani, M. Reiffers, in
{\it Extremely high density of magnetic excitations at T=0 in YbCu$_{5-x}$Au$_x$}, Phys.
Rev. B {\bf 90} (2014) 224409.

\bibitem{YbPd2In} F. Gastaldo, A. Dzubinsk\`a, M. Reiffers, G. Prist\`as, I. Curl\`ik,
J.G. Sereni, M. Giovannini, in {\it YbPd$_ 2$In: a promising candidate to strong entropy
accumulation at very low temperature}, ArXiv [cond-mat] 1711.02335, 27 June 2018.

\bibitem{Gegenw} Y. Tokiwa, B. Piening, H.S. Jeevan, S.L. Bud�ko, P.C. Canfield, P.
Gegenwart,in {\it Super-heavy electron material as metallic refrigerant for adiabatic
demagnetization cooling}, Sci. Adv. 2 (2016) e1600835.

\bibitem{YbCu4Ni} J.G. Sereni, I. Curl�k, M. Giovannini, A. Strydom, M. Reiffers, in {\it
Physical properties of the very heavy fermion YbCu4Ni}, arXiv:1805.08051v1
[cond-mat.str-el] 21 May 2018

\bibitem{YbPt2Sn} T. Gruner, D. Jang, A. Steppke, M. Brando, F. Ritter, C. Krellner, C.
Geibel, in {\it Unusual weak magnetic exchange in YbPt$_2$Sn and YbPt$_2$In}, J. Phys.:
Condens. Matter {\bf 26} 485002 (2014).

\bibitem{JangNatur} D. Jang, T. Gruner, A. Steppke, K. Mistsumoto, C. Geibel, M. Brando,
in {\it Large magnetocaloric effect and adiabatic demagnetization refrigeration with
YbPt$_2$Sn}, Nature Communications, ncomms9680 (2015).

\bibitem{JLTP18} J.G. Sereni, in {\it Entropy constraints in the ground state formation of
magnetically frustrated systems}, J. Low Temp. Phys. {\bf 190} (2018) 1-19, and references
therein.

\bibitem{Stewart01} G.R. Stewart, in {\it Non-Fermi-liquid behavior in d- and f-electron
metals}, Rew. Mod. Phys. {\bf 73} (2001) 797.

\bibitem{SerPhysB18} J.G. Sereni, in {\it Role of the entropy in the ground state
formation of magnetically frustrated systems}, Physica B: Condensed Matter {\bf 536}
(2018) 397.

\bibitem{Ni9Cox} L. Peyker, C. Gold, W. Scherer, H. Micho, E-W. Scheidt, in {\it Competing
magnetic interactions in CeNi$_{9-x}$Co$_x$Ge$_4$}, J. of Phys.: Conf. Series {\bf 273}
(2011) 012049.

\bibitem{Ni9Cux} L. Peyker, C. Gold, E-W. Scheidt, W. Scherer, J.G. Donath, P. Gegenwart,
F. Mayr, T. Unruh, V. Eyert, E. Bauer, H Michor, in {\it Evolution of quantum criticality
in CeNi$_{9-x}$Cu$_x$Ge$_4$}, J. of Phys.: Cond. Mat. {\bf 21} (2009) 235604.

\bibitem{CMN2} R. A. Fisher, E. W. Hornung, G. E. Brodale, and W. F. Giauque, in {\it
Magnetothermodynamics of Ce$_2$Mg$_3$(N0$_3$)$_{12}$.24H$_2$O}, J. Chem. Phys. {\bf } 58
(1973) 5584.

\bibitem{Ce4} N. Kurita, H. Lee, Y. Tokiwa, C. F. Miclea, E. D. Bauer, F. Ronning, J. D.
Thompson, Z. Fisk, P. Ho, M. B. Maple, P. Sengupta, I. Vekhter, R. Movshovich, in {\it
Thermal and magnetic properties of the low-temperature antiferromagnet
Ce$_4$Pt$_{12}$Sn$_{25}$}, Phys. Rev. B {\bf 82} (2010) 174426.

\bibitem{Ser07} J.G. Sereni, in {\it Peculiar thermal features of Ce-systems around their
critical points} J. Low Temp. Phys. {\bf 147} (2007) 179.

\bibitem{Ramirez06} R. Moessner and A.P. Ramirez, in {\it Geometrical Frustration};
Physics Today, February 2006, p.24. \bibitem{YbCo2Zn20} M.S. Torikachvili, S. Jia, E.D.
Mun, S.T. Hannahs, R.C. Black, W.K. Neils, D. Martien, S.L. Bud'ko, P.C. Canfield, in {\it
Six closely related related YbT$_2$Zn$_{20}$ heavy fermion compounds with large local
moment degeneracy}, PNAS {\bf 104} (2007) 9960.

\bibitem{Sereni15} J.G. Sereni, in {\it Entropy Bottlenecks at $T\to 0$ in Ce-Lattice and
Related Compounds}, J Low Temp Phys {\bf 179} (2015) 126.

\bibitem{Pippard} A.B. Pippard, in {\it Elements of classical Thermodynamics}, University
Press, Cambridge, 1964. \bibitem{YbSn3Ru4} T. Klimczuk, C.H. Wang, J.M. Lawrence, Q. Xu,
T. Durakiewicz, F. Ronning, A. Llobet, F. Trouw, N. Kurita, Y. Tokiwa, Han-oh Lee, C.H.
Booth, J.S. Gardner, E. D. Bauer, J.J. Joyce, H.W. Zandbergen, R. Movshovich, R.J. Cava,
J.D. Thompson, in {\it Crystal fields, disorder, and antiferromagnetic short-range order
in} Yb$_{0.24}$Sn$_{0.76}$Ru, Phys. Rev. B {\bf 84} (2011) 075152.

\bibitem{Elsevier} J.G. Sereni, in: {\it Magnetic Systems: Specific Heat}, Saleem Hashmi
(editor-in-chief), Materials Science and Materials Engineering. Oxford: Elsevier; 2016.
pp. 1-13; ISBN: 978-0-12-803581-8.

\end{thebibliography}
\end{document}